\documentclass[aps,prb,showpacs,twocolumn,superscriptaddress]{revtex4}
\usepackage{amsmath,graphics}
\usepackage[next]{inputenc}
\usepackage[dvips]{epsfig}
\def\be{\begin{equation}}
\def\ee{\end{equation}}

\def\bi{\begin{itemize}}
\def\ei{\end{itemize}}
\def\bn{\begin{enumerate}}
\def\en{\end{enumerate}}
\def\bea{\begin{eqnarray}}
\def\eea{\end{eqnarray}}
\def\no{\nonumber}
\def\ba{\begin{array}}
\def\ea{\end{array}}
\def\bd{\begin{displaymath}}
\def\ed{\end{displaymath}}

\begin{document}
\title{Renormalization of concurrence:\\ the application of quantum renormalization group
to the quantum information systems}

\author{M. Kargarian}
\affiliation{Physics Department, Sharif University of Technology, Tehran 11155-9161, Iran}
\author{R. Jafari}
\affiliation{Institute for Advanced Studies in Basic Sciences, Zanjan 45195-1159, Iran}
\affiliation{Institute for Studies in Theoretical Physics and Mathematics, Tehran 19395-5531, Iran}
\author{A. Langari}
\affiliation{Physics Department, Sharif University of Technology, Tehran 11155-9161, Iran}
\affiliation{Institute for Studies in Theoretical Physics and Mathematics, Tehran 19395-5531, Iran}
\email[]{langari@sharif.edu}
\homepage[]{http://spin.cscm.ir}

\begin{abstract}
We have combined the idea of renormalization group and quantum
information theory. We have shown how the entanglement or
concurrence evolve as the size of the system being large, i.e.
the finite size scaling is obtained. Moreover, It introduces how the
renormalization group approach can be implemented to obtain the
quantum information properties of a many body system. We have
obtained the concurrence as a measure of entanglement, its
derivatives and their scaling behavior versus the size of system for
the one dimensional Ising model in transverse field. We have found
that the derivative of concurrence between two blocks each
containing half of the system size diverges at the critical point
with the exponent which is directly associated 
with the divergence of the correlation length.
\end{abstract}
\date{\today}


\maketitle
A fundamental difference between quantum and classical physics is
the possible existence of nonclassical correlations in quantum
systems called Entanglement\cite{Bell}.
 Recently, the study of
strongly correlated systems in condensed matter physics from the
perspective notions of quantum information theory  has been received
much attentions. It seems that the main motivations for such treatment
are two folds: (i) Over the last decade the
entanglement has been realized to be a crucial resource to process
and send information in novel ways such as quantum teleportation,
supercoding and algorithms for quantum computations\cite{Nielsen},
(ii) the novel features of the ground state of many body systems
which consists of a superposition of huge number of product states
opens the question of how this states are interrelated.

The role of entanglement in quantum phase transition (QPT)
\cite{Sachdev} is of considerable interest\cite{Osterloh}. Quantum
phase transitions occur at absolute zero
and are driven
by quantum fluctuations.
Entanglement as a direct measure of quantum correlations shows
nonanalytic behavior such as discontinuity in the vicinity of the
quantum phase transition point \cite{Wu}. In the past few years the
subject of several activities  were to investigate the behavior of
entanglement in the vicinity of quantum critical point for different
spin models \cite{Osterloh,Vidal1,Vidal2,Bose,Osborne,Verstraete} as
well as itinerant systems \cite{Zanardi,Gu,Anfossi}.

Our main purpose in this work is to combine the idea of quantum
renormalization group \cite{pfeuty, wilson} and quantum information
theory. This will give two insights on (i) how a quantum information
property (QIP) evolves as the size of system becomes large and
(ii) QRG connects the nonanalytic behavior of entanglement to the
critical phenomenon properties of the model.
To have a concrete discussion, the one dimensional
$S=\frac{1}{2}$ Ising model in a transverse field (ITF) has been
considered by implementing the quantum renormalization group (QRG)
approach.


The main idea of the RG method is the mode elimination or thinning
of the degrees of freedom followed by an iteration which reduces
the number of variables step by step until reaching a fixed point.
In Kadanoff's approach, the lattice is divided into blocks.
Each block is treated independently to build the projection
operator onto the lower energy subspace. The projection of the
inter-block interaction is mapped to an effective Hamiltonian ($H^{eff}$)
which acts on the renormalized subspace \cite{miguel1,langari}.


We have considered the ITF model on a periodic chain of $N$ sites
with the Hamiltonian 
\be
H=-J\sum_{i=1}^{N}(\sigma_{i}^{z}\sigma_{i+1}^{z}+g\sigma_{i}^{x}).
\label{eq8} \ee 
To implement QRG the Hamiltonian is divided to
two-site blocks \cite{qg-miguel}, $H^{B}=\sum_{I=1}^{N/2}h_{I}^{B}$ with
$h_{I}^{B}=-J(\sigma_{1,I}^{z}\sigma_{2,I}^{z}+g\sigma_{1,I}^{x})$.
The remaining part of the Hamiltonian is included in the inter-block
part,
$H^{BB}=-J\sum_{I=1}^{N/2}(\sigma_{2,I}^{z}\sigma_{1,I+1}^{z}+g\sigma_{2,I}^{x})$,
where $\sigma_{j,I}^{\alpha}$ refers to the $\alpha$-component of
the Pauli matrix at site $j$ of the block labeled by $I$. The
Hamiltonian of each block ($h_{I}^{B}$) is diagonalized exactly
and the projection operator ($P_{0}$) is constructed from the two
lowest eigenstates,
$P_{0}=|\psi_{0}\rangle\langle\psi_{0}|+|\psi_{1}\rangle\langle\psi_{1}|$.
In this respect the effective Hamiltonian
($H^{eff}=P_{0}[H^{B}+H^{BB}]P_{0}$) is similar to the original one
(Eq.(\ref{eq8})) replacing the couplings with the following
renormalized coupling constants. \bea \label{eq16}
J'=J\frac{2(\sqrt{g^{2}+1}+g)}{1+(\sqrt{g^{2}+1}+g)^{2}}~~,~~
g'=g^{2}\eea


Since the block Hamiltonian is treated exactly, the density matrix
can be written in terms of the ground state of the two site-block,
$\rho_{12}=|\psi_{0}\rangle\langle\psi_{0}|$.
We confine our interest to the entanglement between two sites which is
measured by concurrence. The concurrence in terms
of the parameters defined for a two site-block is:
\bea \label{concurrence}
C=Max\{\lambda_{1}-\lambda_{2}-\lambda_{3}-\lambda_{4},0\} ,
\label{eq21} \eea where $\lambda_{k}(k=1,2,3,4)$ are the square
roots of the eigenvalues in descending order of the operator
$R_{12}$: \bea \no R_{12}=\rho_{12}\tilde{\rho}_{12}~~~,~~~
\tilde{\rho}_{12}=
(\sigma^{y}_{1}\otimes\sigma^{y}_{2})\rho_{12}^{\ast}(\sigma^{y}_{1}\otimes\sigma^{y}_{2}).
\label{eq22} \eea

The concurrence as a
measure of entanglement is a local quantity which includes the
global properties of a system. Generally, the global properties of a
system enters into the entanglement effectively by summing over the
whole degrees of freedom except the local one. In other words, a
system can be supposed of a single site and a heat bath (the rest of
system). It is supposed that the effect of a heat bath can be
replaced by an effective single site quantity, {\it the
entanglement}. The effective single site represents the long range
properties of the model and not the microscopic ones. Having this in
mind we can enter the global properties of the model to the
entanglement (the local quantity) using the renormalization group
idea. In this respect, we always think of a two site model which can
be treated exactly as obtained above. However, the coupling
constants of the two site model are the effective ones which are
given by the renormalization group procedure. This can be used as an
new approach to calculate the entanglement in a large system.

\begin{figure}
\begin{center}
\includegraphics[width=8cm]{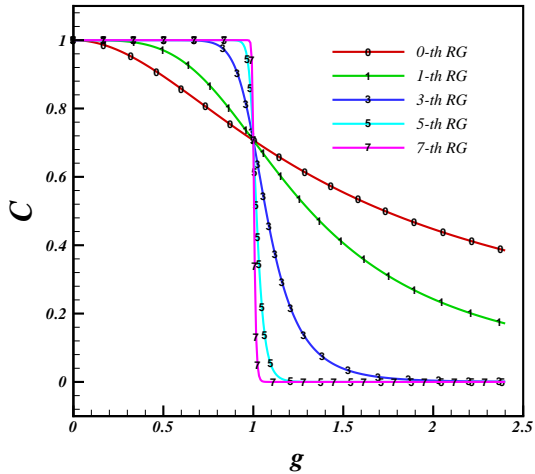}
\caption{(color online) Concurrence of the ITF chain as a function of g
(transverse field strength normalized to the exchange interaction)
at different RG steps.} \label{fig2}
\end{center}
\end{figure}

\begin{figure}
\begin{center}
\includegraphics[width=8cm]{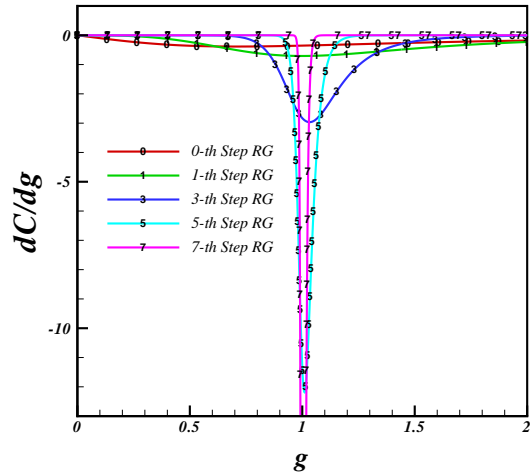}
\caption{(color online)Evolution of the first derivative Concurrence
under RG. in the limit of large system (high RG step), the
nonanalyticity behavior of the first derivative of concurrence is
captured through the diverging.} \label{fig3}
\end{center}
\end{figure}


Before implementing the idea of renormalization group to calculate
the entanglement, let first briefly analyze the RG equations
(Eq.(\ref{eq16})). The RG equations give the flow of coupling
constants in the phase diagram. Any point in the phase diagram runs
to a stable fixed point (after enough iterations) which defines the
stable phases. The unstable fixed points are important because they
define the critical points which are at the border of different
stable phases. 
There are two stable and one unstable fixed point for the RG
equations of ITF model. The stable phases are represented by two
fixed points $g=0$ (long ranged ordered Ising phase) and $g=\infty$
(the paramagnetic phase). The critical point $g_{c}=1$ is the
transition point between the Ising ($g<g_{c}$) and paramagnetic
($g>g_{c}$) phases. The obtained critical point is exactly the same as
what can be obtained by transforming the ITF model to
free fermions using the Jordan-Wigner transformation \cite{Latorre}.

To implement the idea of RG approach in concurrence we have plotted
in Fig.\ref{fig2} the value of $C$ versus $g$ for different RG
steps. In the the n-th RG step the expression given in
Eq.(\ref{eq21}) is evaluated at the renormalized coupling given by
the n-th iteration of $g$ given in Eq.(\ref{eq16}). The zeroth RG
step means a bare two site model while in the first RG step the
effective two site model represents a four site chain. Generally, in
the n-th RG step, a chain of $2^{n+1}$ sites is represented
effectively by the two sites with renormalized couplings. All plots
in Fig.\ref{fig2} cross each other at the critical point, $g_{c}=1$.
In other words the critical point is a scale free point where the
quantum fluctuations extend over all length scales. 

After few RG steps (which represent an enough large system) the
concurrence at the critical point ($g_{c}$) of the model is
discontinues that is a signature of phase transition. As shown in
Fig.\ref{fig2}, the concurrence switches  from one to zero. It means
that the entanglement acts like an order parameter, i.e for the
paramagnetic phase the entanglement is zero and for the long ranged
ferromagnetic phase it is equal to one.
The non-analytic behavior in some physical quantity is a feature of
second-order quantum phase transition. It is also accompanied by a
scaling behavior since the correlation length diverges and there is
no characteristic length scale in the system at the critical point.
Osterloh,\emph{et. al}\cite{Osterloh} have verified that the
entanglement in the vicinity of critical point of ITF model and XY
model in transverse field shows a scaling behavior. They have
concentrated on the concurrence between the two nearest-neighbor
sites at various sizes of system. As we have stated in the RG
approach for ITF model, a large system, i.e. $N=2^{n+1}$, can be
effectively described by two sites with the renormalized couplings
of the n-th RG step. Thus, the entanglement between the two
renormalized sites represents the entanglement between two parts of
the system each containing $N/2$ sites effectively. In this respect
we can speak of {\it block entanglement} -the entanglement between a
block and the rest of system- in a large system provided the size of
the block and the rest of system is equal.

\begin{figure}
\begin{center}
\includegraphics[width=8cm]{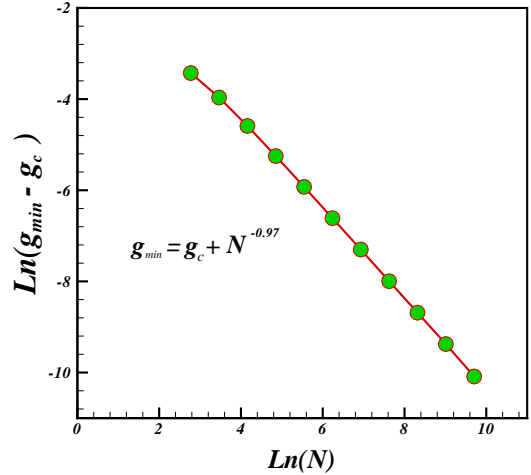}
\caption{(color online) Scaling of the position ($g_{m}$) of the
minimum of concurrence at different RG steps, it is seen that
$g_{m}$ goes to $g_{c}$ as the size of the system becomes large as
$g_{m}=g_{c}+N^{-0.97}$.} \label{fig4}
\end{center}
\end{figure}

\begin{figure}
\begin{center}
\includegraphics[width=8cm]{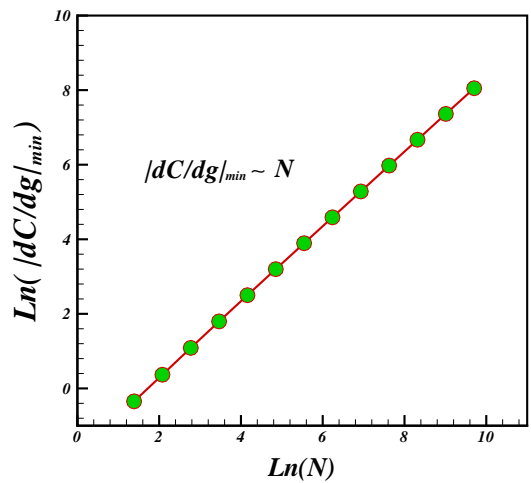}
\caption{(color online) Scaling of the minimum of the first
derivative of concurrence for various sizes of system. The RG
procedure shows the minimum diverges as
$\mid\frac{dC}{dg}|_{g_m}\mid\sim N$. } \label{fig6}
\end{center}
\end{figure}

Having this in mind, the first derivative of concurrence is analyzed
as a function of coupling $g$ at different RG steps which manifest
the size of the system. The derivative of concurrence with respect
to the coupling constant ($\frac{dC}{dg}$) shows a singular behavior
at the critical point (Fig.\ref{fig3}). 
The singular behavior is the result of discontinuous change of $C$
at $g=g_c$. We have plotted $\frac{dC}{dg}$ versus $g$ in
Fig.\ref{fig3} for different RG steps which shows the singular
behavior as the size of system becomes large (higher RG steps). A
more detailed analysis shows that the position of the minimum
($g_m$) of $\frac{dC}{dg}$ tends towards the critical point like
$g_{m}=g_{c}+N^{-\theta}$ with $\theta=0.97$, which has been plotted in
Fig.\ref{fig4}. A similar behavior has been reported in Osterloh's
work \cite{Osterloh} which shows that $\lambda_{m}$ scales as
$\lambda_{m}=1+N^{-1.87}$. It should be noticed that $\lambda_{m}$
is the position of the minimum of the derivative of concurrence of
two nearest neighbor sites which is different from our case. In our
treatment the derivative of the concurrence of two blocks shows a
singular behavior. Moreover, we have derived the scaling behavior of
$y\equiv|\frac{dC}{dg}|_{g_m}|$ versus $N$. This has been plotted in
Fig.\ref{fig6} which shows a linear behavior of $ln(y)$ versus
$ln(N)$. The scaling behavior is
$\mid\frac{dC}{dg}|_{g_m}\mid\sim N^{\theta}$ with exponent $\theta=1$. However, the
minimum value of derivative of concurrence for two nearest neighbor
sites diverges logarithmically \cite{Osterloh}, $\frac{dC}{d
\lambda}|_{\lambda_m}=-0.2702 \ln N$. 

\begin{figure}
\begin{center}
\includegraphics[width=8cm]{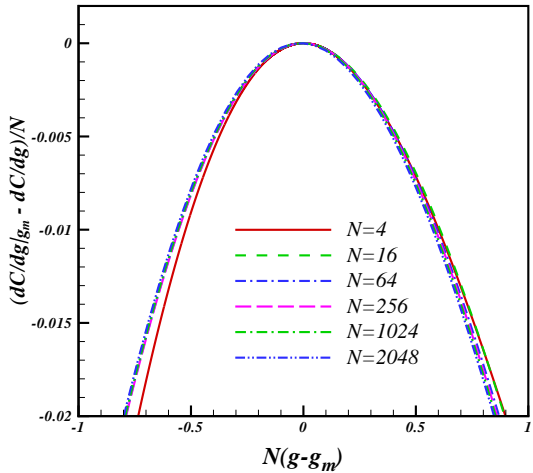}
\caption{(color online) a manifestation of finite size scaling
through RG treatment. The curves corresponding to different lattice
sizes collapse to a single graph. } \label{fig5}
\end{center}
\end{figure}

We would like to show that the exponent $\theta$ is directly related to
the correlation length exponent close to the critical point. The
correlation length exponent, $\nu$, gives the behavior of
correlation length in the vicinity of $g_{c}$, 
i.e $\xi \sim (g-g_{c})^{-\nu}$. Under the RG transformation Eq.(\ref{eq16}) the
correlation length scales  in the $n^{th}$ RG step as $\xi^{(n)} \sim
(g_{n}-g_{c})^{-\nu}=\xi/n_{B}^{n}$ which immediately leads to an
expression for $\mid\frac{dg_{n}}{dg}\mid_{g_{c}}$ in terms of $\nu$
and $n_{B}$ (number of sites in each block). 
Dividing the last equation to $\xi \sim (g-g_{c})^{-\nu}$
gives $\mid\frac{dg_{n}}{dg}\mid_{g_{c}}\sim N^{1/\nu}$ which implies 
$\theta=1/\nu$, since $\mid\frac{dC}{dg}|_{g_m}\sim \mid\frac{dg_{n}}{dg}\mid_{g_{c}}$ 
at the critical point.
It should also be noted that
the scaling of the position of minimum , $g_{min}$
(Fig.\ref{fig4}), also comes from the behavior of the correlation
length near the critical point. As the critical point is approached
and in the limit of large system size, 
the correlation length almost covers the size of the system,
i.e. $\xi \sim N $, and a simple comparison with $\xi \sim (g-g_{c})^{-\nu}$
results in the following scaling form $g_{m}=g_{c}+N^{-1/\nu}$.

To get the finite size scaling behavior of concurrence 
we have followed a scaling trick in which all graphs
collapse on each other. This is also a manifestation of the existence of
finite size scaling for the block entanglement. The
scaling trick is based on the divergence of
derivative of concurrence close to the critical point
(Fig.\ref{fig3}) and the power law scaling obtained in Figs.(\ref{fig4}, \ref{fig6}). 
We define the scaling variable $x=N^{1/\nu}(g-g_{m})$ and obtain the
universal function $F(x)$ such that $\frac{dC}{dg} \sim N^{1/\nu}F(x)$ where $F(x)=1/(1+x^{2})$.
In this way we have plotted
$(\frac{dC}{dg}|_{g_{m}}-\frac{dC}{dg})/N$ versus $N(g-g_{m})$ in
Fig.\ref{fig5}. The curves which correspond to different system
sizes clearly collapse on a single universal curve. This results
justify that the RG implementation of entanglement  truly capture
the critical behavior of the ITF model.

To summarize, we have initiated the idea of renormalization group
(RG) to study the quantum information properties (QIP) of a system.
In this respect some basic notions have been introduced: (i) The
evolution of QIP, i.e. entanglement or concurrence, in terms of RG
steps give how this properties develop from a finite size system to
its thermodynamic counterpart. In other words, there exist some
finite size scaling for this properties. (ii) The RG procedure can
be implemented to obtain QIP of a system in terms of effective
Hamiltonian which is described by renormalized coupling constants.
(iii)The RG procedure indicates that the exponents governing the
nonanalytic behavior of the QIP in the vicinity of the critical
point comes from the long range properties of the model. This is the
manifestation of the fact that some global properties of the system
can be represented by local properties such as entanglement. This
notions have been observed and approved in our study of ITF model.
Moreover, RG approach show that as the size of the system becomes
large two effective sites are entangled for ferromagnetic , i.e.
$g<g_{c}$ and disentangled for paramagnetic , i.e.$g>g_{c}$, and
probably manifest the block entanglement is  relevant for the
existence of ferromagnetic correlations.

We have also implemented the idea of QRG to the anisotropic Heisenberg model (XXZ) \cite{kargarian}
to get the QIP.
We have used the Von-Neuman entropy for searching the entanglement between
one site and rest of a three site block.
We have been able to get the scaling
behavior of the concurrence and entanglement of this model.
This justifies our main idea.

The atuthors would like to thank M. A. Martin-Delgado for fruitful discussions and
comments. This work was supported in part by the Center of Excellence in Complex
Systems and Condensed Matter (www.cscm.ir).

\end{document}